\documentclass[english,journal]{IEEEtran}
\usepackage[T1]{fontenc}
\usepackage[latin1]{inputenc}

\newtheorem{lemma}{Lemma}

\usepackage{amsfonts,latexsym,graphicx}
\usepackage{hyperref}

\usepackage{babel}

\begin{document}

\title{New $[48,16,16]$ Optimal Linear Binary Block Code}

\author{Axel Kohnert%
\thanks{The author is with the Mathematical Department, University of Bayreuth,
D-95440 Bayreuth, Germany%
}}
\maketitle
\begin{abstract}
A new $[48,16,16]$ optimal linear binary block code is given. To
get this code a general construction is used which is also described
in this paper. The construction of this new code settles an conjecture
mentioned in paper by Janosov et al. \cite{MR2446769}  where the
authors found an new optimal $[47,15,16]-$code, which is relevant
to the applied construction.\end{abstract}

\section{Introduction}

\PARstart{A} linear binary $[n,k]$-code is a $k$-dimensional subspaces
of the $n$-dimensional vector space $GF(2)^{n}$ over the finite
field $GF(2)$ with $2$ elements. The $2^{k}$ codewords of length
$n$ are the elements of the subspace, they are written as row vectors.
The weight $wt(c)$ of a codeword is defined to be the number of nonzero
entries of $c$ and the minimum distance $dist(C)$ of a code $C$
is the minimum of all weights of the nonzero codewords in $C.$ For
the purpose of error correcting codes we are interested in codes with
high minimum distance $d$ as these allow the correction of $\left\lfloor (d-1)/2\right\rfloor $
errors. An $[n,k]-$code with minimum distance $d$ is called an $[n,k,d]-$code.
On the other hand we are interested in codes of small length $n$.
High minimum distance and small length are contrary goals for the
optimization of codes. There are several \cite{grassl-online-table,brouwer-online-table}
tables giving upper limits for the minimum distance of a linear code
of fixed length $n$. A linear code $C$ is called optimal if $dist(C)$
is equal to this known upper bound. In a recent paper the authors
first found by a computer program and later constructed an new optimal
$[47,15,16]-$code. They also conjectured, that there is an $[48,16,16]-$code.
As $16$ is the maximum possible minimum distance of a code with parameters
$n=48$ and $k=16$ such will be also optimal. In this paper we give
a construction of such a code.

\section{A different $[47,15,16]-$code}

Using the methods described in \cite{micbra2,BraunKohnertWassermann:05a}
we constructed like the authors of \cite{MR2446769} also an optimal
$[47,15,16]-$code. A key ingredient of the methods described in our
earlier papers is the prescription of some automorphisms of the linear
code we want to construct. In the case of the $[47,15,16]-$code we
prescribed a cyclic group $G$ of order $10$ generated by the matrix\[
\left(\begin{array}{ccccccccccccccc}
1 & 0 & 1 & 0 & 0 & 1 & 0 & 1 & 0 & 0 & 1 & 1 & 1 & 1 & 1\\
1 & 0 & 1 & 0 & 0 & 0 & 0 & 0 & 0 & 0 & 1 & 0 & 1 & 1 & 1\\
1 & 1 & 1 & 1 & 1 & 1 & 1 & 1 & 1 & 0 & 0 & 1 & 0 & 0 & 0\\
1 & 1 & 1 & 0 & 0 & 0 & 1 & 0 & 1 & 0 & 1 & 1 & 0 & 0 & 0\\
1 & 1 & 0 & 1 & 0 & 1 & 0 & 1 & 1 & 1 & 1 & 0 & 1 & 0 & 0\\
1 & 1 & 1 & 0 & 0 & 1 & 0 & 0 & 0 & 0 & 1 & 1 & 1 & 1 & 1\\
0 & 1 & 1 & 0 & 1 & 1 & 0 & 1 & 0 & 1 & 0 & 1 & 0 & 0 & 0\\
1 & 0 & 0 & 0 & 1 & 1 & 0 & 0 & 1 & 1 & 0 & 0 & 0 & 1 & 0\\
1 & 1 & 1 & 0 & 1 & 1 & 0 & 1 & 0 & 1 & 0 & 0 & 1 & 1 & 1\\
1 & 0 & 0 & 1 & 1 & 1 & 0 & 0 & 1 & 1 & 0 & 0 & 1 & 1 & 1\\
0 & 1 & 1 & 1 & 1 & 0 & 1 & 1 & 1 & 1 & 1 & 0 & 1 & 1 & 0\\
1 & 0 & 1 & 0 & 1 & 1 & 0 & 0 & 0 & 1 & 1 & 1 & 1 & 0 & 1\\
1 & 0 & 1 & 1 & 0 & 0 & 0 & 1 & 0 & 0 & 0 & 0 & 0 & 0 & 0\\
1 & 0 & 1 & 1 & 1 & 0 & 0 & 0 & 1 & 0 & 1 & 0 & 1 & 1 & 1\\
1 & 1 & 1 & 0 & 1 & 1 & 0 & 0 & 0 & 0 & 1 & 0 & 0 & 0 & 1\end{array}\right).\]
Using this group we get a Diophantine system of equations with $3383$
variables. These are the number of orbits of $G$ on the possible
columns of a generator matrix of a $15-$dimensional binary linear
code. We found a solution of this system by the heuristic methods
described in \cite{MR2506408,zwanzger-heuristics-2007} and could
construct a generator matrix of an optimal $[47,15,16]-$code by combining
$7$ orbits. The generator matrix of the code found by this method
is as follows:

~

{\footnotesize 00000111110000011111000001111100011000110100000 }{\footnotesize \par}

{\footnotesize 00011000111111100111001110000100111001110000011 }{\footnotesize \par}

{\footnotesize 00101000010001100001010000111100011000111000000 }{\footnotesize \par}

{\footnotesize 01101011110010101011010000001101001011000101111 }{\footnotesize \par}

{\footnotesize 01010101110000011101100010000001010000000000011 }{\footnotesize \par}

{\footnotesize 00111110011100101000010101110101001001010011101 }{\footnotesize \par}

{\footnotesize 01010100110100100111111101000010100101000110111 }{\footnotesize \par}

{\footnotesize 11001110110001010001011100010110110111111010100 }{\footnotesize \par}

{\footnotesize 10111000001001111111101100010011101000110010010 }{\footnotesize \par}

{\footnotesize 11011110011001110100111011001101100001101010001 }{\footnotesize \par}

{\footnotesize 01010111101010000110111011001001101010110011011 }{\footnotesize \par}

{\footnotesize 00110011000100000100010111001110111010100000011 }{\footnotesize \par}

{\footnotesize 01001101001000000100111111000011111111000010111 }{\footnotesize \par}

{\footnotesize 00110001111100001011010001101111010111001001100 }{\footnotesize \par}

{\footnotesize 01110110110000101001110000101100101101110101010}{\footnotesize \par}

~

The weight enumerator of this code is: \[
\begin{array}{c}
1+1082x^{16}+2560x^{18}+3360x^{20}+6656x^{22}+9000x^{24}\\
+5632x^{26}+2400x^{28}+1536x^{30}+541x^{32}\end{array}.\]

The maximum weight of this code is $32$, which also shows that this
code is non-isomorphic to the code found in \cite{MR2446769}. So
this is a second optimal $[47,15,16]-$code.

\section{The $[48,16,16]-$code}

The 'advantage' of this new $[47,15,16]-$code $C$ is the low maximum
weight, which is only $32$ compared to the maximum weight $36$ of
the code with the same parameters in \cite{MR2446769}. The low maximum
weight allows the following two-step construction of a new optimal
$[48,16,16]-$code. We first append a zero column to the generator
matrix $\Gamma$ of the code $C.$ This gives a generator matrix $\Gamma'$
a $[48,15,16]-$code $C'$ with maximum weight $32.$ Now we add the
all-one codeword (it has weight $48$ and therefore it is not in $C'$)
as a further row to the increased generator matrix $\Gamma'$ giving
a $[48,16]-$code $\hat{C.}$ The effect of the all-one codeword is
that we can write $\hat{C}$ as the disjoint union of $C'$ and the
complement $\overline{C'}$. And as the weight of the complement of
a codeword $v$ is $n-wt(v)$ we know that the minimum weight of the
codewords in $\overline{C'}$ is $48-32=16$ as the maximum weight
in $C'$ was $32.$ This shows that the minimum distance of $\hat{C}$
is $16,$ and we constructed this way an new optimal $[48,16,16]-$code.
Using the description $\hat{C}=C'\cup\overline{C'}$ we also get the
complete weight enumerator of the new code which is:\[
\begin{array}{c}
1+1623x^{16}+4096x^{18}+5760x^{20}+12288x^{22}+18000x^{24}\\
+12288x^{26}+5760x^{28}+4096x^{30}+1623x^{32}+x^{48}.\\
\\\end{array}\]

This construction works in general in all cases where we know the
maximum weight of the binary linear code. It also works in cases where
we append more than $1$ zero column to the original generator matrix.
This is summarized in the following:
\begin{lemma}
Let $C$ be a binary $[n,k,d]-$code with generator matrix $\Gamma$
and maximum weight $d'.$ Let $\hat{C}$ be the linear $[n+p,k+1]-$code
whose generator matrix we get by first appending $p$ zero columns
to $\Gamma$ and then finally an all-one row as the last row to the
increased generator matrix. Then the minimum distance of $\hat{C}$
is \[
min\{d,n+p-d'\}.\]

\end{lemma}

\section{Conclusion}

As summarized in the above lemma we can apply this construction in
all cases where we know the maximum weight. To apply this method for
the construction of an new $[n,k+1,d]-$code it is necessary to find
an $[n-p,k,d]-$code with maximum weight $\le n-d.$ So the problem
is to construct codes with prescribed minimum and maximum weight.
But it is possible to modify the methods described in \cite{micbra2,MR2506408,zwanzger-heuristics-2007,BraunKohnertWassermann:05a,BraunKohnertWassermann:05c}
(which originally only take the minimum weight into account) to cover
this case and it is hoped that we can find using this modified approach
together with the above mentioned method further improvements of the
international tables of best linear codes.

\bibliographystyle{plain}
\bibliography{codes}

\end{document}